\documentclass[aps,pre,twocolumn,showpacs,superscriptaddress,groupedaddress,10pt]{revtex4}

\usepackage{graphicx}
\usepackage{color} 
\usepackage{epstopdf}
\usepackage{epsfig}
\usepackage{amssymb}
\usepackage{amsmath}
\usepackage{setspace}
\usepackage{float}
\begin{document}

\title{Formation of adhesion domains in stressed and confined membranes}

\author{Nadiv Dharan$^1$ and Oded Farago$^{1,2}$\\
$^1$Department of Biomedical Engineering, $^2$Ilse Katz Institute\\
for Nanoscale Science and Technology, Ben Gurion University,\\
Be'er Sheva 84105, Israel}

\begin{abstract}

The adhesion bonds connecting a lipid bilayer to an underlying surface
may undergo a condensation transition resulting from an interplay
between a short range attractive potential between them, and a long
range fluctuation-induced potential of mean force. Here, we use
computer simulations of a coarse-grained molecular model of supported
lipid bilayers to study this transition in confined membranes, and in
membranes subjected to a non-vanishing surface tension. Our results
show that confinement may alter significantly the condensation
transition of the adhesion bonds, whereas the application of surface
tension has a very minor effect on it. We also investigate domain
formation in membranes under negative tension which, in free
membranes, causes enhancement of the amplitude of the membrane thermal
undulations. Our results indicate that in supported membranes, this
effect of a negative surface tension on the fluctuation spectrum is
largely eliminated by the pressure resulting from the mixing entropy
of the adhesion bonds.

\end{abstract}
\maketitle

\vspace{0.45cm}

\newpage

\section{Introduction}
\label{sec:intro}

Lipid bilayers serve as a physical barrier separating the content of
the cell from its surroundings. They are embedded with different types
of proteins responsible for numerous biological functions. An
important class of membrane proteins are cell adhesion molecules that
form specific (ligand-receptor) bonds with different biological
elements, e.g., the extracellular matrix, the cytoskeleton, and
neighboring cell membranes~\cite{alberts}. Cell adhesion plays a
central role in many biological processes, such as T-cell activation
as part of the immune response~\cite{springer}, cell
migration~\cite{Huttenlocher} and tissue formation~\cite{Vleminckx}.

When membranes adhere to another surface, the adhesion bonds may
condense to form large adhesion
domains~\cite{wang,monks,Selhuber-Unkel}, which not only provide
mechanical stability but also promote cellular signaling pathways
necessary for many biological processes~\cite{Burridge,Giancotti}. The
aggregation of adhesion bonds is driven by attractive forces,
including electrostatic and Van der Walls forces~\cite{Israelachvili},
interactions resulting from cytoskeleton remodeling~\cite{Wulfing},
and membrane-mediated interactions~\cite{Bruinsma}. The latter
mechanism stems from the suppression of membrane thermal fluctuation
caused by the adhesion. The associated decrease in fluctuation entropy
is minimized when the bonds cluster together into a single adhesion
domain.

Over the past two decades, considerable efforts have been made to
better understand the role of membrane mediated interactions in the
formation of adhesion domains. To this end, a common theoretical
approach is the use of discrete lattice models, where the occupied
sites represent membrane adhesion bonds, while empty sites stand for
free fluctuating membrane
segments~\cite{Lipowsky,Lipowsky2,Weikl,Speck,Speck2,Smith,WF}. Computer
simulations of such models have shown that membrane fluctuations can
facilitate conditions for condensation of adhesion
bonds. Nevertheless, it has been also concluded that the fluctuation
induced interactions alone are too weak to promote aggregation and,
hence, other attractive interactions between the bonds must also
exist. A similar conclusion was recently reached in simulation studies
employing coarse-grained (CG) molecular models, which provide a more
physical description of the system. In one such study, we simulated a
bilayer membrane with a small fraction of lipids from the lower
monolayer connected to an underlying surface, and additionally
introduced a short range attractive potential of depth $\epsilon$
between the adhesive lipids~\cite{dharan}. Our simulations revealed
that the adhesive lipids underwent a first order condensation
transition when the strength of the short range potential between them
exceeds a threshold value $\epsilon_c>0$. The fact that adhesion
domains do not form for $\epsilon=0$ implies that the fluctuation
entropy gained by the aggregation of the adhesive lipids only
partially compensates for their loss of mixing entropy. In another CG
simulation study, Noguchi demonstrated that cell junctions connecting
{\em more than two}\/ membranes can aggregate without an additional
attractive potential between them~\cite{Noguchi}. This observation can
be understood because the total entropy of thermal undulations
increases linearly with the number of membranes, $N_m$, while the
mixing entropy of the cell junctions is independent of $N_m$.

Previous simulation studies have focused on the aggregation behavior
of adhesion bonds in {\em tensionless} membranes. Here, we wish extend
our investigations to lipid bilayers subjected to surface tension and
confinement. Under such conditions, the membrane long wavelength
thermal undulations are suppressed, which weakens the fluctuation
induced attraction. Specifically, a harmonic confining potential of
strength $\gamma>0$ suppresses thermal fluctuations at length scales
larger than $\xi_\gamma\sim(\kappa/\gamma)^{1/4}$, where $\kappa$ is
the bending modulus of the membrane. Likewise, surface tension
$\tau>0$ affects bending modes with wavelengths greater than
$\xi_\tau\sim(\kappa/\tau)^{1/2}$. The influence of tension and
confinement on the fluctuation mediated potential of mean force (PMF)
was analyzed for the case of two adhesion
bonds~\cite{Farago_PRE81}. It was shown that the PMF depends
logarithmically on the pair distance: $\phi(r)=Ck_BT\ln(r)$, where
$C=2$ at short distances, $C=1$ at long distances ($r\gg\xi_{\tau}$)
for stressed membranes, and $C=0$ for long distances
($r\gg\xi_{\gamma}$) for confined membranes. These findings suggest
that tension and confinement may hinder the formation of adhesion
domains, and that the phase transition into the condensed state
requires a stronger short range attraction between the
bonds. Therefore, one should expect the transition threshold value
$\epsilon_c$ to grow with $\tau$ and $\gamma$; however, the magnitude
of the increase in $\epsilon_c$ remains hard to assess due to the many
body nature of the PMF~\cite{WF}. One may also consider the case of a
bilayer membrane under negative tension. Negative tensions amplify the
amplitude of the long wavelength undulations (and, for very large
negative values, even lead to instability), and this should cause a
reduction in $\epsilon_c$. It is especially interesting to examine
whether $\epsilon_c$ can decrease to zero, in which case the adhesion
domains will form without an additional short range potential, i.e.,
on purely entropic grounds. As will be shown further on, the results
of the current study suggest that confinement might have a very strong
impact on $\epsilon_c$, while the application of a positive surface
tension hardly influences the transition into an aggregated phase. In
cases where the membrane is subjected to a negative surface tension,
we observe the formation of elongated adhesion domains close to the
transition point.

\section{Monte Carlo Simulations}
\label{sec:MC}

In order to address the above issues, we conducted Monte Carlo
simulations of a lipid bilayer using the Cooke-Deserno
implicit-solvent CG model~\cite{Deserno}. The details of the
simulations can be found in ref.~\cite{dharan}. Briefly, we simulated
bilayers of $2N=2000$ lipids (1000 lipids per monolayer), where each
lipid consists of one hydrophilic (head) bead and two hydrophobic
(tail) beads of size $\sigma$.  A flat impenetrable surface was placed
below the membrane at $z=0$, and adhesion was established by
connecting $N_b$ head beads from the lower monolayer to it. A
Lennard-Jones type potential with a tunable depth $\epsilon$ was
introduced between the adhesive head beads. We simulated systems with
various densities of adhesive lipids, $\phi=N_b/N$, and for different
values of the interaction potential depth $\epsilon$. Our simulations
for confined membranes included an additional impermeable surface
placed above the upper monolayer at $z=z_{\rm conf}$. Simulations of
membranes under constant mechanical surface tension $\tau$ were
carried out according to the method described in
ref.~\cite{Farago_gene2}.

\begin{figure}[t]
\begin{center}%0.275
%\scalebox{0.4}{\centering\includegraphics{fig1.eps}}
{\centering\includegraphics[width=0.5\textwidth]{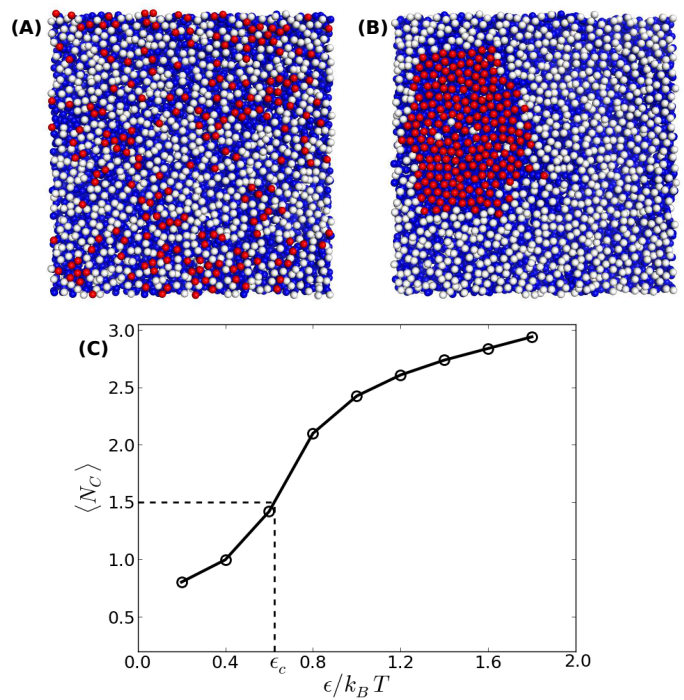}}
\end{center}
\vspace{-0.5cm}
\caption{Bottom view of equilibrium snapshots of a tensionless bilayer
  with $\phi=0.2$ showing (A) a scattered distribution of adhesion
  bonds for $\epsilon=0.2k_BT$, and (B) a condensed phase for
  $\epsilon=1.4k_BT$. The head and tail beads are colored in white and
  blue, respectively, while the adhesive beads are colored in red. (C)
  The average number of contacts between the adhesion bonds as
  function of $\epsilon$. The condensation transition value
  $\epsilon_c$ is empirically defined via the equality $\langle
  N_C\rangle\simeq1.5$. The line serves as a guide to the reader's
  eye.}
\label{fig:fig1}
\end{figure}

Typical equilibrium configurations of tensionless membranes with
$\phi=0.2$ are shown in Fig.~\ref{fig:fig1} for $\epsilon=0.2k_BT$ (A)
and $\epsilon=1.4k_BT$ (B). In the former, the adhesive lipids (whose
head beads are shown in the figure in red) are scattered across the
surface, while in the latter they condense into a large
cluster. Fig.~\ref{fig:fig1}(C) shows the average energy $E$ of pair
interactions between adhesive beads, normalized per bond and expressed
in dimensionless units by dividing by the potential strength
$\epsilon$. This quantity, which is a measure for the typical number
of contacts between the adhesive beads $\langle N_C\rangle$, increases
sharply when the condensation transition takes place. From the results
of Fig.~\ref{fig:fig1}(C), we empirically determine the transition
value as the one for which $\langle N_C\rangle\simeq1.5$.

\section{Results}
\label{sec:results}
\subsection{Membranes under confinement}
\label{sec:conf}

\begin{figure}[t]
\begin{center}%0.275
\scalebox{0.33}{\centering\includegraphics{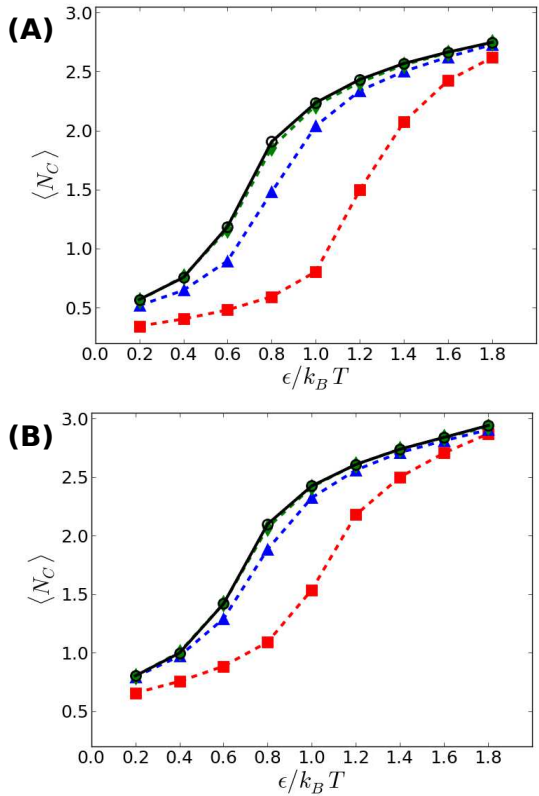}}
%{\centering\includegraphics[width=0.5\textwidth]{fig2.eps}}
\end{center}
\vspace{-0.5cm}
\caption{The average number of contacts between the adhesion bonds in
  membranes confined by a surface located at $z_{\rm conf}=9\sigma$
  (green diamonds), $z_{\rm conf}=7.5\sigma$ (blue triangles) and
  $z_{\rm conf}=6\sigma$ (red squares), as a function of $\epsilon$,
  for (A) $\phi=0.1$, and (B) $\phi=0.2$. Results for non-confined
  membranes are denoted by black circles. The lines serve as a guide
  to the reader's eye. The statistical errors are comparable to the
  size of the symbols.}
\label{fig:fig2}
\end{figure}

We first start with our simulation results for non stressed membranes
confined between the supporting underlying surface (on which the
adhesive beads reside) and a second upper surface. The former is
located at $z=0$, just underneath the tips of the head beads of the
lower leaflet, while the latter is placed at $z=z_{\rm conf}\geq
6\sigma$. The degree of confinement increases when $z_{\rm conf}$
decreases which, in turn, would lead to stronger suppression of the
membrane thermal fluctuations and a shift in the transition threshold
$\epsilon_c$ to larger values. This trend is demonstrated in
Fig.~\ref{fig:fig2}, showing our simulation results for $z_{\rm
conf}=6\sigma$, $7.5\sigma$, and $9\sigma$ in supported membranes with
$\phi=0.1$ (A) and $\phi=0.2$ (B). For $z_{\rm conf}=9\sigma$, our
results (green diamonds) match perfectly with the results obtained for
non-confined membranes (black circles). This means that the rate of
collisions between the membrane and the upper surface is negligibly
small and, therefore, it does not affect the fluctuation
spectrum. Lowering the confining surface by a distance equal to the
size of a bead and a half to $z_{\rm conf}=7.5\sigma$ has a more
noticeable effect on membrane thermal undulations, which leads to an
increase in the condensation transition value from
$\epsilon_c\simeq0.65k_BT$ for non-confined membranes at $\phi=0.1$ to
$\epsilon_c\simeq 0.8k_BT$. For $\phi=0.2$, the shift is smaller, from
$\epsilon_c\simeq 0.6k_BT$ to $\epsilon_c\simeq0.7k_BT$. When the
upper surface is further lowered to $z_{\rm conf}=6\sigma$, it touches
the tips of the head beads in the upper leaflet, as the thickness of
the bilayer is equal to the size of six beads. A confining surface
located at $z_{\rm conf}=6\sigma$ completely suppresses thermal
undulations, and eliminates the fluctuation mediated interactions
between the adhesion bonds. Under these conditions, the threshold
value for aggregation increases to $\epsilon_c\simeq1.2k_BT$ at
$\phi=0.1$, and $\epsilon_c\simeq 1k_BT$ for $\phi=0.2$. These
values are approximately twice larger than the corresponding values
found when no upper plate exists ($\epsilon_c\simeq0.65k_BT$ and
$\epsilon_c\simeq0.6k_BT$, respectively), which is in accord with the
conclusions of refs.~\cite{WF,Noguchi}, that the entropic gain of
aggregation compensates for, roughly, half of the loss in mixing
entropy of the adhesion bonds.

\subsection{Membranes under surface tension}
\label{sec:surf}
\begin{figure}[t]
\begin{center}
\scalebox{0.33}{\centering\includegraphics{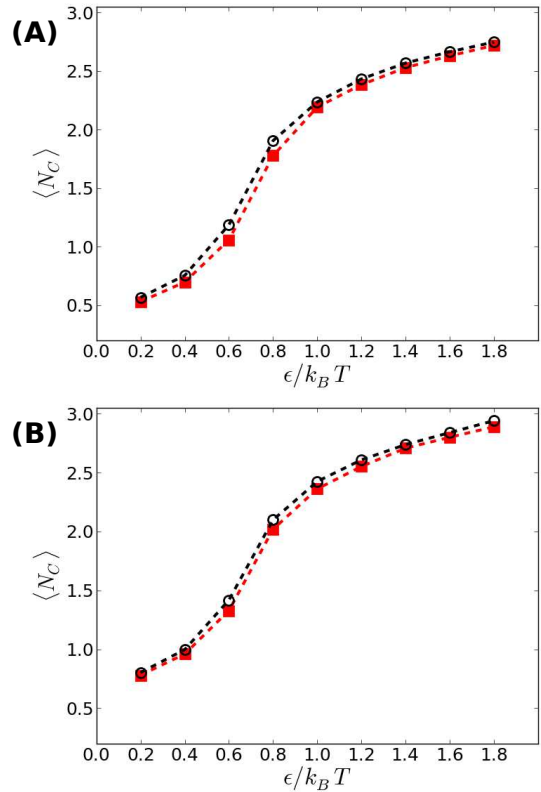}}
%{\centering\includegraphics[width=0.5\textwidth]{fig3.eps}}
\end{center}
\vspace{-0.5cm}
\caption{The average number of contacts between the adhesion bonds in
membranes under mechanical surface tension $\tau=0.48 k_BT/\sigma^2$,
as a function of $\epsilon$ for (A) $\phi=0.1$, and (B)
$\phi=0.2$. The results are plotted in red squares, and are compared
with the results for tensionless membranes ($\tau=0$) that are
depicted by black open circles. The lines serve as guides to the
eye. The statistical errors are comparable to the size of the
symbols.}
\label{fig:fig3}
\end{figure}

\begin{figure}[t]
\begin{center}
\scalebox{0.33}{\centering\includegraphics{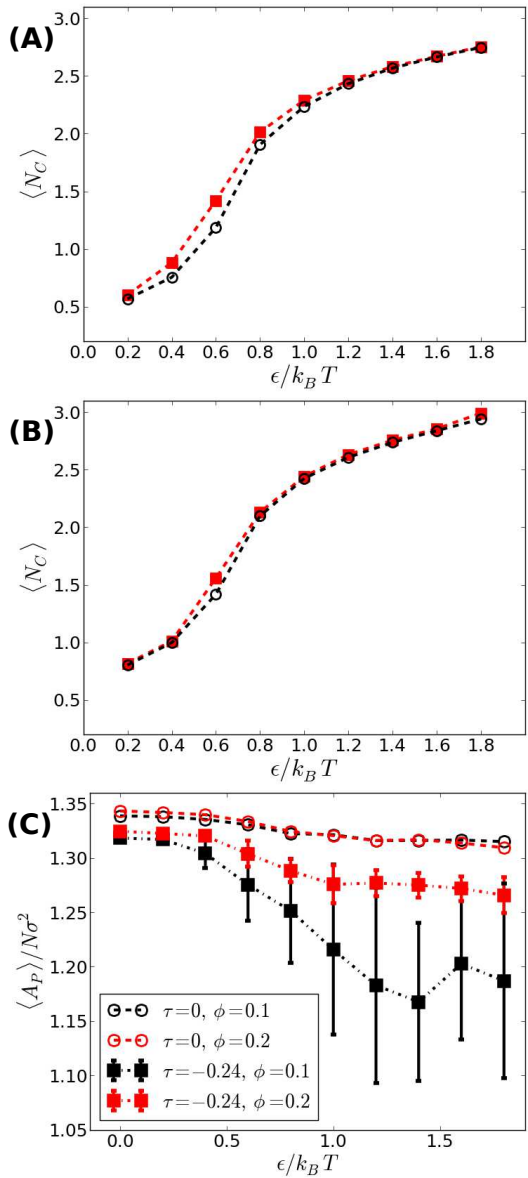}}
%{\centering\includegraphics[width=0.5\textwidth]{fig4.eps}}
\end{center}
\vspace{-0.5cm}
\caption{The average number of contacts between the adhesion bonds in
  membranes under surface tension $\tau=-0.24 k_BT/\sigma^2$, as a
  function of $\epsilon$ for (A) $\phi=0.1$, and (B) $\phi=0.2$. The
  results are plotted in red squares, and are compared with the
  results for tensionless membranes ($\tau=0$) that are depicted by
  black open circles. (C) The mean projected area per lipid as a
  function of $\epsilon$, for $\phi=0.1$ (black) and $\phi=0.2$
  (red). Circles denote the results for tensionless membranes, while
  the results for $\tau=-0.24k_BT/\sigma^2$ are shown in squares. The
  lines serve as guides to the eye.}
\label{fig:fig4}
\end{figure}

We now proceed to our simulation results for non-confined membranes
subjected to a lateral surface tension $\tau>0$. Since, just like
steric confinement, surface tension also suppresses the thermal
undulations of the membrane, one may expect to also find an increase
in the value of $\epsilon_c$. However, our computational results that
are depicted in Fig~\ref{fig:fig3}, demonstrate that the $\langle
N_C\rangle$ vs.~$\epsilon$ curve remains almost unchanged when the
tension increases from $\tau=0$ (black open circles) to
$\tau=0.48k_BT/\sigma^2$ (red squares) for both $\phi=0.1$ (A) and
$\phi=0.2$ (B). This observation suggests that the fluctuation
spectrum of the simulated membranes is very little affected by the
tension, which is expected given that the bending modulus of the model
membrane has been previously estimated to be
$\kappa\simeq8k_BT$~\cite{Farago_mode}. For $\tau=0.48k_BT/\sigma^2$,
the wavelength of the modes that are suppressed are larger than
$\xi_{\tau}=2\pi\sqrt{\kappa/\sigma}\simeq25\sigma$, which is much
larger than the typical distance $\bar{d}=\sqrt{A_P/N_b}\simeq4\sigma$
between the adhesion bonds when the latter are uniformly distributed
on the surface (where $A_P$ is the projected area of the membrane).
In order for the effect of tension to be appreciable, one must apply
stronger tensions; however, the application of a stronger tension
would open pores in the membrane and tear it apart. In simulations
with a surface tension of $\tau=0.64k_BT/\sigma^2$, we observed pores
in most of the simulated membranes, and in the few that had managed to
survive the tension, we still did not observe any considerable changes
in the strength of membrane mediated interactions.

We next aimed to address the implications of applying a negative
surface tension. In general, a negative tension imposed on the
membrane leads to a reduction in its projected area, $A_P$, and
amplifies the long wavelength bending modes. Hence, the
fluctuation-induced attraction between the adhesion bonds is expected
to be stronger which, in turn, implies that the threshold for
condensation $\epsilon_c$ should become smaller than in tensionless
membranes. To test this hypothesis, we simulated the membrane under a
negative tension of $\tau=-0.24k_BT/\sigma^2$. We performed two sets
of independent MC simulations, one starting from a random distribution
of adhesion bonds, and another where initially the adhesion bonds were
organized in one large cluster. The system was equillibrated until
configurations originating from these two distinct initial conditions
achieved similar characteristics. Fig.~\ref{fig:fig4} shows our
results for $\langle N_C\rangle$ as a function of $\epsilon$ for
$\phi=0.1$ (A) and $\phi=0.2$ (B). Contrary to our expectation to
observe a reduction in $\epsilon_c$, the data for
$\tau=-0.24k_BT/\sigma^2$ appears almost identical to the results of
the tensionless case, with $\epsilon_c\simeq 0.6k_BT$ for both values
of $\phi$. Taken together with the data for $\tau=+0.48k_BT/\sigma^2$
(see Fig.~\ref{fig:fig3}), we conclude that application of surface
tension has a negligible effect on the condensation transition, within
the range of tensions where the supported membrane is mechanically
stable. 

\begin{figure}[t]
\begin{center}
%scalebox{0.125}{\centering\includegraphics{./Figs/fig5_new2.eps}}
\centering\includegraphics[width=0.465\textwidth]{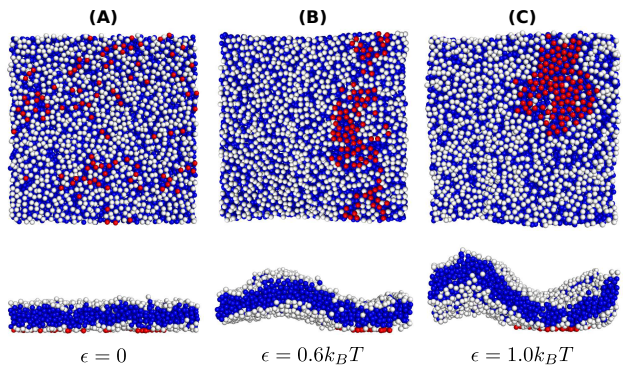}

\end{center}
\vspace{-0.5cm}
\caption{Typical equilibrium configurations of membranes under surface
  tension $\tau=-0.24k_BT/\sigma^2$ with density of adhesion bonds
  $\phi=0.1$ for (A) $\epsilon=0$, (B) $\epsilon=0.6k_BT$ and (C)
  $\epsilon=1.0k_BT$. The figures in the upper and lower rows display
  bottom and side views of the system, respectively. Color coding is
  the same as in Fig.~\ref{fig:fig1}.}
\label{fig:fig5}
\end{figure}

The negative tension, however, does have an impact on the shape of
membranes. Freely fluctuating bilayers assume buckled configurations
at negative tensions larger (in absolute value) than
$\tau_c\simeq-4\pi^2\kappa/A_P$~\cite{Noguchi2,Otter}. In this study,
$A_P/N\simeq1.33\sigma^2$ [see Fig.~\ref{fig:fig4}(C)], which gives
$\tau_c\simeq-0.24k_BT/\sigma^2$. In ref.~\cite{yotam}, a similar
model membrane consisting of the same number of lipids was simulated
and, indeed, for $\tau=\tau_c$ the membrane appeared quite buckled. In
supported membranes, however, the emergence of buckled configurations
occurs only in membranes with large adhesion domains (i.e., for
$\epsilon\gtrsim\epsilon_c$). Fig.~\ref{fig:fig5} shows typical
equilibrium configuration for $\phi=0.1$ with $\epsilon=0$ (A),
$0.6k_BT$ (B), and $1.0k_BT$ (C). Each configuration is shown both in
side and bottom views. When the attractive potential is set to
$\epsilon=0$, the distribution of the adhesion bonds is scattered and
the membrane remains fairly flat. This indicates that the mixing
entropy of the bonds dominates the fluctuation entropy of the bilayer,
despite the imposed negative tension [see Fig.~\ref{fig:fig5}(A)]. For
$\epsilon=1.0k_BT$, the short range pair interactions between the
bonds lead to their aggregation. Once the bonds condense, their
influence on the thermal behavior of the membrane is greatly weakened,
and strong bending undulations appear [see
  Fig.~\ref{fig:fig5}(C)]. Close to the condensation transition, at
$\epsilon=0.6k_BT$, the system exhibits some interesting features: The
amplitude of one of the two longest wavelength bending modes [with
  wavevector $\vec{q}_{(1,0)}=(2\pi/\sqrt{A_P})(1,0)$, or
  $\vec{q}_{(0,1)}=(2\pi/\sqrt{A_P})(0,1)$] grows considerably, and
the membrane assumes an anisotropic buckled configuration. The
adhesion bonds are concentrated throughout the minimum of the
dominating bending mode, forming an elongated domain (``stripe'') [see
  Fig.~\ref{fig:fig5}(B)]. These observed characteristics represent an
intricate balance between the driving forces that govern the
thermodynamic behavior of the system. Under negative tension, the
system benefits from a reduction in the projected area, leading to a
decrease in the Gibbs free energy. The membrane, however, is quite
incompressible and, thus, the reduction in $A_P$ must be accompanied
by an increase in the area stored in thermal fluctuations whose
amplitudes grow. The modes that experience the largest increase in
amplitude are the softest ones, corresponding to $\vec{q}_{(1,0)}$ and
$\vec{q}_{(0,1)}$~\cite{Noguchi2} (see footnote ref.~\cite{foot2}). In
our simulations, we seldom observed situations where both these modes
were simultaneously excited~\cite{foot1}, which can be linked to the
mixing entropy of the adhesion bonds. When only one of the long modes
is dominant, the contact area between the membrane and the surface,
which is available for the presence of the adhesion bonds, is larger
than in configurations where both modes are excited.

\begin{figure}[t]
\begin{center}
%\scalebox{0.225}{\centering\includegraphics{fig6.eps}}
{\centering\includegraphics[width=0.5\textwidth]{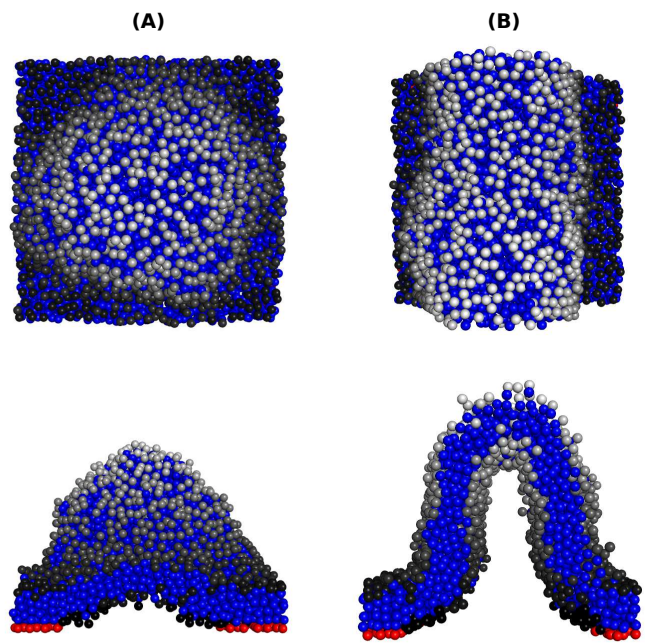}}
\end{center}
\vspace{-0.5cm}
\caption{Configurations of membranes with $\phi=0.1$ and
  $\epsilon=1.0k_BT$ under a strong negative tension
  $\tau=-0.32k_BT/\sigma^2$, showing a spherical protrusion (A) and a
  tubular one (B). The upper and lower rows display top and side views
  of the membrane respectively. The tail and adhesive beads are
  colored in blue and red, respectively. The head beads are colored in
  grayscale to reflect their height above the surface, with lighter
  colors representing a higher bead.}
\label{fig:fig6}
\end{figure}

The interplay between the mixing entropy of the adhesion bonds and the
contribution of the negative tension to the free energy is further
demonstrated in Fig.~\ref{fig:fig4}(C), depicting the mean projected
area $\langle A_p \rangle$ (normalized by the number of lipids per
monolayer $N=1000$) as a function of $\epsilon$ for
$\tau=-0.24k_BT/\sigma^2$ (squares) and $\tau=0$ (circles). In the
tensionless case, we observe a very mild decrease in $\langle
A_P\rangle$ as $\epsilon$ increases, occurring mainly around
$\epsilon_c$. For $\tau=-0.24k_BT/\sigma^2$, $\langle A_P\rangle$
maintains a value close to the tensionless case for
$\epsilon<\epsilon_c$, and drops significantly for
$\epsilon>\epsilon_c$. In the latter regime, we also observe an
increase in the area fluctuations, resulting in larger uncertainties
(error bars) in our estimates of $\langle A_p\rangle$. The sharp
decrease in $\langle A_P\rangle$ and the concurrent increase in the
area fluctuations are anticipated outcomes of a negative surface
tension~\cite{yotam}. The fact that they are observed only above the
condensation transition is consistent with the picture discussed in
the previous paragraph that, below $\epsilon_c$, the effect of the
negative tension is largely eliminated by the pressure resulting from
the mixing entropy of the adhesion bonds. Notice that the sharp
decrease in the mean projected area and the increase in the area
fluctuations are much more noticeable for $\phi=0.1$ than for
$\phi=0.2$. This is to be expected because the smaller $\phi$, the
smaller the restrictions imposed by the adhesion bonds on large
thermal undulations which are directly coupled to the projected area
by the highly incompressible character of the membrane.

Applying an even stronger negative tension causes the membrane to lose
its mechanical stability. This is demonstrated in Fig.~\ref{fig:fig6},
showing snapshots in top and side views of membranes with $\phi=0.1$
and $\epsilon=1.0k_BT$ subjected to $\tau=-0.32k_BT/\sigma^2$. In
these snapshots, the head beads are colored in grayscale, with lighter
colors indicating beads located higher above the underlying
surface. The application of a strong negative tension causes the
supported membrane to develop large protrusions with either spherical
[Fig.~\ref{fig:fig6}(A)] of tubular [Fig.~\ref{fig:fig6}(B)]
shapes. The adhesion bonds (which are colored in red, and are only
partially visible) are concentrated in the periphery of the
protrusion, where the membrane is in contact with the underlying
surface. We note that the observed protrusions tend to evolve slowly
and, therefore, the snapshots shown in Fig.~\ref{fig:fig6} may not
represent true equilibrium structures. On the other hand, we also note
that very similar equilibrium structures, featuring spherical and
tubular protrusions, have been recently observed in an experimental
study where supported lipid bilayers were subjected to
compression~\cite{Staykova}.

\section{Summary}
\label{sec:summary}

The thermal fluctuations of a supported membrane induce an attractive
PMF between the adhesion bonds, that promotes the formation of
adhesion domains. In previous studies of systems where the adhesion
bonds interact with each other via a pairwise short range attractive
potential, it has been concluded that the fluctuation-induced PMF
significantly decreases the condensation transition point,
$\epsilon_c$, of the adhesion bonds. In this work, we investigate
adhesion domain formation in stressed and confined membranes by using
MC simulations of a CG supported bilayer model. Our simulations reveal
that placing a plate above the membrane (in addition to the surface
supporting the membrane from below), significantly alters the
condensation point. In the most extreme case, where the membrane is
completely flattened between the two plates, the value of $\epsilon_c$
increases by almost a factor of two. In contrast to the impact of
steric confinement, the application of a positive surface tension,
which also acts to suppress membrane undulations, hardly affects the
transition within the range of tensions where the supported membrane
is mechanically stable. Similarly, we find that a negative tension
also has a very minor effect on the condensation
transition. Nevertheless, once the adhesion bonds are aggregated into
a large domain (i.e., for $\epsilon>\epsilon_c$), the negative tension
affects the shape of the membrane and causes it to buckle. Below
$\epsilon_c$ the adhesion bonds are scattered across the membrane,
which prevents the formation of buckled configurations. From the fact
that $\epsilon_c$ is hardly affected by the application of a negative
tension, we conclude that the mixing entropy of the adhesion bonds
dominates the demixing entropy of thermal fluctuations. Close to
$\epsilon_c$, we observed the formation of elongated adhesion
stripes. Such configurations emerge from an interplay between the
mixing entropy of the adhesion bonds, the short range residual
potential, and the applied negative tension. Finally, under a very
strong negative tension, we observe tubular and spherical structures
protruding out of the membrane's plane, which indicates that the
system is at the onset of mechanical instability.

This work was supported by the Israel Science Foundation (ISF) through
grant No.~1087/13.


\begin{thebibliography}{99}
\bibitem{foot2} Notice that in ref.~{\protect \cite{Noguchi2}}, the
  membrane is simulated in the fixed-area ensemble, $A_P=L_xL_y$, with
  $L_x\ne L_y$ which generates buckled configurations with anistropic
  surface tension. In this study, we simulate the fixed (isotropic)
  tension ensemble, where the projected area $A_P$ is allowed to
  fluctuate, but with $L_x=L_y$.
\bibitem{foot1} Out of 15 independent realizations of the system
  corresponding to $\epsilon=0.6k_BT$, in more than half we observed
  one dominant long Fourier mode. For larger values of $\epsilon$,
  realizations where both modes were simultaneously excited occurred
  more frequently.
\end{thebibliography}
\end{document}